\documentclass[doublecol]{epl2}
\usepackage[]{graphicx}
\usepackage{amsmath}
\usepackage{amssymb}
\newcommand{\vct}[1]{\mathbf{#1}}

\renewcommand\Re{\operatorname{Re}}
\renewcommand\Im{\operatorname{Im}}

\title{Non-equilibrium Casimir forces: Spheres and sphere-plate}
\shorttitle{Non-equilibrium Casimir forces: Spheres and sphere-plate}
\date{\today}

\author{Matthias Kr\"uger\inst{1} \and Thorsten Emig\inst{2} \and Giuseppe Bimonte\inst{3} \and Mehran Kardar\inst{1}}
\shortauthor{M. Kr\"uger, T. Emig, G. Bimonte and M. Kardar}

\institute{ 
  \inst{1} Massachusetts Institute of Technology, Department of
  Physics, Cambridge, Massachusetts 02139, USA\\
  \inst{2} Laboratoire de Physique Th\'eorique et Mod\`eles
  Statistiques, CNRS UMR 8626, B\^at.~100, Universit\'e Paris-Sud, 91405
  Orsay cedex, France\\
\inst{3} Dipartimento di Scienze Fisiche, Universit{\`a} di Napoli Federico II, Complesso Universitario MSA, Via Cintia, I-80126 Napoli, Italy and INFN Sezione di Napoli, I-80126 Napoli, Italy 
}
\abstract{ We discuss non-equilibrium extensions of the Casimir force (due to electromagnetic fluctuations), where the objects as well as the environment are held at different temperatures.  While the formalism we develop is quite general, we focus on a sphere in front of a plate, as well as two spheres, when the radius is small compared to separation and thermal wavelengths. In this limit the forces can be expressed analytically in terms of the lowest order multipoles, and corroborated with results obtained by diluting parallel plates of vanishing thickness.  Non-equilibrium forces are generally stronger than their equilibrium counterpart, and may oscillate with separation (at a scale set by material resonances).  For both geometries we obtain stable points of zero net force, while two spheres may have equal forces in magnitude {\it and} direction resulting in a self-propelling state.  }


\begin{document}
\maketitle

The original quantum Casimir effect \cite{Casimir48} is due to zero point fluctuations of the electromagnetic (EM) field in the vacuum between perfectly reflecting objects.  Subsequently Lifshitz \cite{Lifshitz56} treated the more realistic case of dielectric media at finite temperature by considering fluctuating currents inside the objects, including both zero point and thermal fluctuations.  In general, the former dominate the force at small separations, while at separations large compared to the thermal wavelength $\lambda_T$, thermal effects prevail \cite{Lifshitz56,Milonni}.  In situations out of equilibrium, the current fluctuations in each body have to be treated separately at the corresponding temperature, e.g., using fluctuational electrodynamics introduced by Rytov over 60 years ago~\cite{Rytov3}.  Recently, out of equilibrium Casimir forces have been computed in a number of cases including parallel plates~\cite{Antezza08}, modulated plates~\cite{Bimonte09}, as well as a plate and an atom in different setups \cite{Henkel02,Antezza05,Ellingsen10}. There also exists a large literature on forces between atoms or molecules in non-equilibrium \cite{Kweon93,Power94,Cohen03,Rodriguez10}.  Formalisms for treating multiple objects at different temperatures have been recently presented~\cite{Messina,Kruger11}.  In particular, for compact objects, radiation from the environment contributes to the force and has to be incorporated.

Here, we treat (analytically as well as numerically) the cases of two spheres and a sphere in front of a plate. Keeping the description as simple and concise as possible, we focus on the regime where the spheres are small compared to the separation (non-equilibrium effects are in most cases negligible at small separations), as well as thermal wavelengths.  These restrictions allow the use of a one reflection approximation, as well as limiting to the spheres' (frequency-dependent) dipole response, respectively. We find a variety of interesting effects: The forces can be repulsive, oscillate or admit stable (zero force) points.  At large separations, non-equilibrium forces decay as $1/d^2$ for two spheres and become independent of distance for sphere and plate.  We also find points in which a pair of spheres experiences forces of equal magnitude in the {\it same} direction.  In the absence of other forces, this leads to a cooperative motion of two identical spheres at constant separation, i.e. a self-propelled state.  There are similarities to studies of atoms in non-equilibrium situations which we shall briefly comment upon.

As presented in Ref.~\cite{Kruger11}, our formalism treats $N$ objects (labeled as $j=1\dots N$) in vacuum, held at constant temperatures $\{T_j\}$, and embedded in an environment at temperature $T_{env}$. 
The conceptual starting point is the EM field radiated by isolated objects, each at its
respective temperature, which is then scattered by all objects while the environment is 
treated as an additional embedding ``object.'' 
Physical quantities are then computed from the correlation function $C^{neq}$ of the electric field $\bf {E}$ at frequency $\omega$ and points $\bf r$ and ${\bf r}'$ (both outside all objects)  \cite{Kruger11},
\begin{align}
  C^{neq}&(T_{env},\{T_j\})\equiv\left\langle \vct{E}(\omega;\vct{r})\otimes \vct{E}^*(\omega;\vct{r}') \right\rangle \notag\\
&=C^{eq}(T_{env})+ \sum_j \left[C_j^{sc}(T_j)- C_j^{sc}(T_{env})\right]\, .\,\label{eq:1}
\end{align}
Equation~\eqref{eq:1} highlights the contribution of the different temperatures to the non-equilibrium correlation: $C^{eq}(T_{env})$ is the equilibrium correlation, i.e., with all temperatures held at $T_{env}$ (and including zero point fluctuations). 
It leads to the equilibrium Casimir force at temperature $T_{env}$ and  is regarded as known. The {\it difference} of $C^{neq}(T_{env},\{T_j\})$ from $C^{eq}(T_{env})$ is due to the {\it deviations} of the object temperatures $T_j$ from $T_{env}$. Although dealing with $N+1$ sources, we have thus only to evaluate the $N$ terms $\{C_j^{sc}(T)\}$, the field correlations sourced by object $j$ and scattered by all objects. In Ref.~\cite{Kruger11}, we showed that $C_j^{sc}(T)$ can be derived by first considering the radiation of the object in isolation,
\begin{equation}
C_j(T_{j}) \equiv a_{T_j}(\omega)\mathbb{G}_j \Im \varepsilon_j \mathbb{G}_j^*,
\end{equation}
where  $a_T(\omega)\equiv \frac{\omega^4 \hbar (4\pi)^2}{c^4}(\exp[\hbar \omega/k_BT]-1)^{-1}$, and $\mathbb{G}_j$  is the Green's function of the object. The QED origin of the force is manifested by the speed of light $c$ and Planck's constant $\hbar$.
$C_j(T_{j})$ is found by integration over the environment sources~\cite{Kruger11,Eckhardt83}, subsequent scatterings lead to 
\begin{eqnarray}
C_j^{sc}(T_j)&=&\mathbb{O}_{j}\,C_j(T_j) \,\mathbb{O}_{j}^\dagger\,,\quad{\rm with}\,\label{eq:ms}\\
\mathbb{O}_{j}&=& (1-\mathbb{G}_0\mathbb{T}_{\bar{j}})\frac{1}{1-\mathbb{G}_0\mathbb{T}_j\mathbb{G}_0\mathbb{T}_{\bar{j}}}\, .\notag
\end{eqnarray}
The multiple scattering operator $\mathbb{O}_{j}$ is expressed in terms of the composite $T$-operator $\mathbb{T}_{\bar{j}}$ describing scattering by the other objects (not $j$), 
as well as  the free Green's function $\mathbb{G}_0$.
For two objects, $\mathbb{T}_{\bar{1}}=\mathbb{T}_{2}$ is the operator of the second object.
The force $\vct{F}$ acting on one of the objects (say object $k$) in this non-equilibrium situation is given by the integration of the Maxwell stress tensor ${\boldsymbol\sigma}$ over a surface $S_k$ enclosing only this object, projected onto the surface outward normal $\vct{n}_k$,
\begin{equation}
\vct{F}^k=\Re\oint_{S_k}  {\boldsymbol \sigma}\cdot \vct{n}_{k} \,dA\, . \label{eq:fin}
\end{equation}
The stress tensor  
is related to the field correlations, since 
\begin{align}
\sigma_{ab}(\vct{r})&=\int \frac{d\omega}{16\pi^3}\left\langle E_a E^*_b+B_a B^*_b-\frac{1}{2}\left(|E|^2+|B|^2\right)\delta_{ab}\right\rangle,\notag
\end{align}
where $a,b= 1,2,3$. 
Note that the sum of forces on all objects does not necessarily vanish, (i.e., there can be a net force on the system), and we must consider the force acting on each object separately. From Eq.~\eqref{eq:1}, ${\bf F}^k$ has the following contributions
\begin{align}
{\bf F}^k(T_{env},\{T_j\})={\bf F}^{k,eq}(T_{env})+\!\sum_j \!\left[{\bf F}^k_{j}(T_j)- {\bf F}^k_{j}(T_{env})\right].\label{eq:forcef}
\end{align}
Here, ${\bf F}^{k,eq}(T_{env})$ is the force in equilibrium, and ${\bf F}^k_{j}(T_j)$ is the force acting on object $k$ due to the sources in object $j$ at temperature $T_j$
(obtained from the stress tensor in Eq.~\eqref{eq:fin} for the field $C_j^{sc}(T_j)$). 
\footnote{We note that 
Ref.~\cite{Messina} performs a different decomposition, involving the equilibrium force at the temperatures of the objects rather than at $T_{env}$ as in Eq.~\eqref{eq:forcef}.}

Let us first consider two spheres of radii $R_j$ ($j=1,2$) with complex dielectric and magnetic permeabilities $\varepsilon_j$ and $\mu_j$, at center-to-center distance $d$ and  temperatures $T_j$, embedded in an environment at temperature $T_{env}$.
We derive the total force ${\bf F}^2$ acting on sphere 2; ${\bf F}^1$ is then found by interchanging indices 1 and 2. In Eq.~\eqref{eq:forcef}, ${\bf F}^2$ has three contributions: The equilibrium force for the two spheres evaluated at the temperature of the environment, a contribution due to the deviation of $T_1$ from $T_{env}$ (${\bf F}_1^2$) and a contribution due to the deviation of $T_2$ from $T_{env}$ (${\bf F}_2^2$). The force ${\bf F}_1^2$ follows from the heat radiation of sphere 1, which can be written in terms of its $T$-operator \cite{Bohren, Kruger11}.  For the case $d\gg R_j$ considered here, a one reflection approximation for the operator in Eq.~\eqref{eq:ms}, $\mathbb{O}_{1}\simeq (1-\mathbb{G}_0\mathbb{T}_2)$, is asymptotically exact. It amounts to a one time scattering of the field radiated by sphere 1 at sphere 2, and subsequently performing the integration in Eq.~\eqref{eq:fin} over a surface enclosing  sphere 2. This integration in terms of spherical waves has been discussed, e.g. in Ref.~\cite{Crichton00}.
The force $\vct{F}_2^2$ is calculated similarly, only here we consider the heat radiation of sphere 2, which is once scattered by sphere 1, and the surface of integration closed around sphere 2. 
Consistent with symmetries, the force ${\bf F}^2$ in Eq.\eqref{eq:forcef} is parallel to the
axis connecting the spheres; we shall denote this component by  $F^2$ and adapt the
notation where a positive sign corresponds to attraction.
The resulting force \cite{Kruger11b}, contains the $T$-operators as well as translation matrices for spherical waves, organized in a series of multipoles of orders $l$. 
To terminate the series at the dipole order ($l=1$), for the equilibrium force between spheres~\cite{Emig07,Zandi10}, it is sufficient to require $d\gg R_j$, while in the non-equilibrium case, we have to additionally require $\lambda_T\gg R_j$ ($\lambda_T=\frac{\hbar c}{k_BT}\approx7.6\mu$m at room temperature). 
This ensures $R_j^*=R_j\omega/c\ll1$ for all relevant frequencies, and we restrict to terms linear in the two $T$-operators ${\cal T}_{j}^{P}\equiv {\cal T}_{j,l=1}^{P}(\omega)$ for polarization $P=N,M$ and $l=1$. Then $F_1^2(T)$ is
\begin{align}
& \lim_{\{d,\lambda_T\}\gg R_j}F_1^2=-\frac{\hbar}{c\pi}\int_0^\infty \frac{\omega \, d\omega}{e^{\frac{\hbar\omega}{k_BT}}-1}\sum_{P,P'}\Re[{\cal T}^P_1]\Biggl[
\frac{9c^2}{\omega^2d^2}\notag\\& \Re[{\cal T}^{P'}_2]+\Im[{\cal T}^{P'}_2]\left(\frac{9c^3}{\omega^3d^3}+\frac{18c^5}{\omega^5d^5}+\frac{81c^7}{\omega^7d^7}\delta_{PP'}\right)
\Biggr]\label{eq:f1}.
\end{align}
In Eq.~\eqref{eq:f1} (and Eqs.~\eqref{eq:selft}, \eqref{eq:spf} and \eqref{eq:Fss} below) we omit terms quadratic in ${\cal T}_{j}^{P}$ for brewity, a simplification which is justified for the cases considered below\footnote{It requires $\{|\Im[{\cal T}_{j}^{P}]|, |\Re[{\cal T}_{j}^{P}]|\}\gg |{\cal T}_{j}^{P}|^2$ in the relevant frequency range. E.g. the emissivity of a sphere contains $-\Re[{\cal T}_{j}^{P}]-|{\cal T}_{j}^{P}|^2$ and vanishes for $|\varepsilon|\to \infty$ or $\Im[\varepsilon]\to0$ where $\Re[{\cal T}_{j}^{P}]\to-|{\cal T}_{j}^{P}|^2$, captured only by inclusion of the quadratic terms\cite{Kruger11b}.}.

For large separations, $F_1^2$ decays as $d^{-2}$ and is repulsive. This originates from momentum transfer to the second sphere via absorption or scattering of photons. The other terms in Eq.~\eqref{eq:f1}, with higher powers in $1/d$, are (in most cases) attractive.
Similarly, the {\it self} force $F_2^2(T)$, reads 
\begin{align}
&\notag\lim_{\{d,\lambda_T\}\gg R_j}F_2^2=\frac{\hbar}{c\pi}\int_0^\infty d\omega\frac{\omega}{e^{\frac{\hbar\omega}{k_BT}}-1}\sum_{P}\Re[{\cal T}_2^P] \\&\Re\Biggl\{\Biggl[({\cal T}^P_1-{\cal T}^{\bar{P}}_1)\left(\frac{9c^2}{\omega^2d^2}+i\frac{27c^3}{\omega^3d^3}\right)-({\cal T}^P_1-\frac{{\cal T}^{\bar{P}}_1}{2})\frac{72c^4}{\omega^4d^4}\nonumber\\&-({\cal T}^P_1-\frac{{\cal T}^{\bar{P}}_1}{8})i\frac{144c^5}{\omega^5d^5}+{\cal T}^P_1\left(\frac{162c^6}{\omega^6d^6}+i\frac{81c^7}{\omega^7d^7}\right)\Biggr]e^{2i\frac{\omega}{c}d}\Biggr\},\label{eq:selft}
\end{align}
originating from radiation of sphere 2.  Here $\bar{P}=M$ if $P=N$ and vice versa. In contrast to $F_1^2$, this term can oscillate as function of $d$ at a scale set by material resonances.  These oscillations originate from interference of two coherent traveling waves from sphere 2: i) a wave going to sphere 1, being reflected back past sphere 2, and ii) a wave emitted in the reverse direction. Depending on $\omega d/c$, one has constructive or destructive interference. As these waves interfere in the exterior region, we expect the oscillations to become weaker as $R_2$ becomes large compared to the penetration (skin) depths. For a sharp resonance of $\varepsilon_2(\omega)$ at $\omega_0$ in Eq.~\eqref{eq:selft}, the oscillations as function of distance have wavelength $\pi c/\omega_0$.

For small insulating spheres of radius $R_j$ (with $\mu_j=1$), we employ the following expansions of the $T$-operator 
\begin{align}
\label{eq:expT}
{\cal T}_{j}^{N}=i\frac{2\omega^3}{3c^3} \alpha_j(\omega) + \mathcal{O}\left({R_j^*}^5\right),\quad
{\cal T}_{j}^{M}=\mathcal{O}\left({R_j^*}^5\right),
\end{align}
in terms of  the complex frequency dependent dipole polarizability, 
\begin{equation}
\alpha_j(\omega)\equiv\frac{\varepsilon_j(\omega)-1}{\varepsilon_j(\omega)+2}R_j^3.\label{eq:pola}
\end{equation}
Higher multipoles ${\cal T}_{j,l}^P$ for $l\geq2$ are of order ${R_j^*}^5$,
and  Eqs.~\eqref{eq:f1} and \eqref{eq:selft} can be simplified by use of Eq.~\eqref{eq:expT}. 
The range of applicability of this approximation depends on material properties. 
An expansion of ${\cal T}_j^P$ in both $R_j^*$ and $\sqrt{\varepsilon}R_j^*$  
shows that the condition $|\sqrt{\varepsilon}|R_j\ll\lambda_T$ (in the relevant frequency range) is sufficient for many materials, including the ones studied below. For $|\varepsilon_j|\gg 1$ (conductors), the expansion is generally not applicable (e.g. $T_j^{M}$ is then of order ${R_j^*}^3$ \cite{Zandi10}).
With Eqs.~\eqref{eq:expT} and \eqref{eq:pola}, one sees that $F_j^2$ is only nonzero if $\Im\varepsilon_j\not=0$ (or for magnetic materials $\Im\mu_j\not=0$), as only lossy spheres emit heat. This holds for any $R_j$.

The leading low temperature behavior of the force for insulators can be derived by requiring $\lambda_T\gg\lambda_0$, where $\lambda_0$ is the wavelength of the lowest resonance of the material. The dielectric functions and polarizabilities are then expanded as \cite{Jackson}
\begin{align}\label{eq:insu}
\varepsilon_j(\omega)&=\varepsilon_{0,j}+i\frac{\lambda_{in,j}\omega}{c}+\mathcal{O}(\omega^2),\\
\alpha_j(\omega)&=\alpha_{0,j} +i\alpha_{i0,j}\frac{\lambda_{in,j}\omega}{c}+\mathcal{O}(\omega^2),\label{eq:al}
\end{align}
with $\varepsilon_{0,j}$, $\lambda_{in,j}$, $\alpha_{0,j}$  and $\alpha_{i0,j}=3R^3_j/(\varepsilon_{0,j}+2)^2$ real. For  $\lambda_T\gg\lambda_0$, the interaction term is then given in closed form,
\begin{align}
&\lim_{\{d,\lambda_T\}\gg R_j}F_1^2=\frac{\hbar c}{3d^2}\frac{\lambda_{in,1}\alpha_{i0,1}}{\lambda_T^7}\Biggl[\frac{-32\pi^7\lambda_{in,2}\alpha_{i0,2}}{5\lambda_T}\notag\\
&+\alpha_{0,2}\left(\frac{32\pi^5\lambda_T}{21d}+\frac{8\pi^3\lambda_T^3}{5d^3}+\frac{18\pi\lambda_T^5}{d^5}\right)\Biggr].\label{eq:FT}
\end{align}
The self force $F_2^2$ does not oscillate to lowest order in temperature and takes a lengthy form \cite{Kruger11b}.  In the limit where $d$ is the largest scale, we have 
\begin{equation}
\lim_{d\gg\lambda_T\gg \{R_j,\lambda_0\}}F_2^2=\frac{60\hbar c}{\pi d^9}\lambda_{in,2}\alpha_{i0,2}\alpha_{0,1}\, . \label{eq:larged}
\end{equation}
While in this range of $d$ the force $F_2^2$ is independent of temperature, it vanishes as $T\to0$ since
with $\lambda_T$ the largest scale ($\lambda_T\gg\{d,R_j,\lambda_0\}$), one has 
\begin{equation}
\lim_{d\gg R_j}F_2^2=\frac{6\pi\hbar c}{d^7\lambda_T^2}\lambda_{in,2}\alpha_{i0,2}\alpha_{0,1},\label{eq:smallT}
\end{equation}
which is identical to $F_1^2$ in this limit, with indices 1 and 2 interchanged.
\begin{figure}
\includegraphics[angle=270,width=0.9\linewidth]{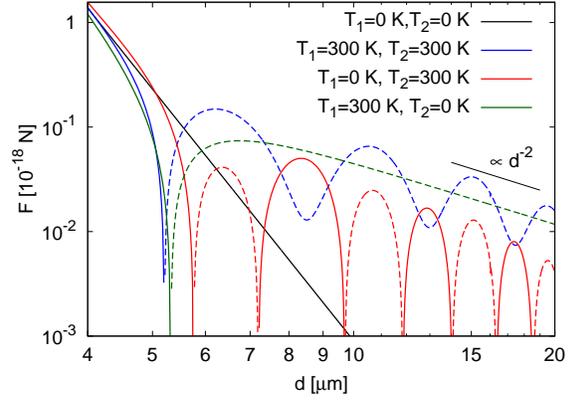}
\caption{\label{fig:1} Total force on sphere 2 in a system of two SiO$_2$ spheres at separation $d$ in a cold (0 K) environment. Dashed lines indicate repulsion. The crossing of solid red and dashed green curves represent a point where the forces are equal in magnitude and direction, see main text. Points of change from repulsive to attractive with increasing $d$ are stable equilibria.} 
\end{figure}
\begin{figure}
\includegraphics[angle=270,width=0.9\linewidth]{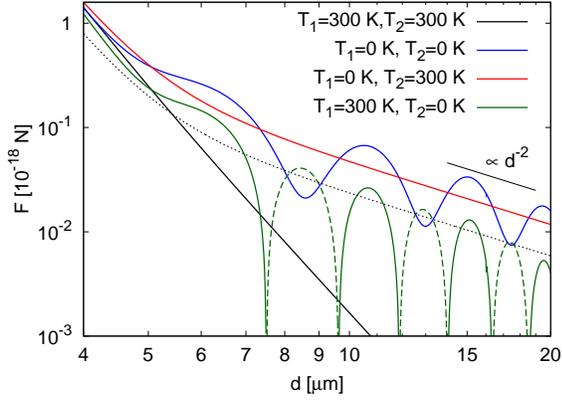}
\caption{\label{fig:1a} 
Total force on sphere 2 in a system of two SiO$_2$ spheres at separation $d$ in a warm ($300 K$) environment. Dashed lines indicate repulsion.  The thin dotted line is the red line divided by 2, see main text.
}
\end{figure}

We evaluate the total force in dipole approximation numerically for $R_1=R_2=1$~$\mu$m using Eq.~\eqref{eq:forcef} and the equilibrium Casimir Polder force (Eq.~(94) in Ref.~\cite{Boyer75}) which in the relevant limits reads
\begin{subequations}\label{eq:feq}
\begin{align}
\lim\limits_{\lambda_T\gg d\gg R_j}F^{2,eq}&=\frac{161}{4\pi}\frac{\hbar c}{d^8}\alpha_{0,1}\alpha_{0,2},\\
\lim\limits_{d\gg \{R_j,\lambda_T\}}F^{2,eq}&=\frac{18\hbar c}{d^7\lambda_T}\alpha_{0,1}\alpha_{0,2}.
\end{align}
\end{subequations}
Figure~\ref{fig:1} shows the forces on SiO$_2$-spheres (we used optical data with $\varepsilon_{0}\approx 3.7$) in a cold (0 K) environment.  We evaluated Eqs.~\eqref{eq:f1} and \eqref{eq:selft} together with \eqref{eq:expT}, \eqref{eq:pola}.  Within these simplifications, the forces are proportional to $R_1^3R_2^3$ ($R=1$~$\mu$m is roughly the upper bound of validity of this approximation for SiO$_2$ at room temperature, where for the total heat emitted by an isolated sphere, the asymptote $\propto R^3$ differs by 12\% from the  exact result \cite{Kruger11}).  
The force starts to deviate strongly from its equilibrium value around $d\approx \lambda_T/2$. Sphere 2 is repelled at large $d$ if $T_1=300 K$ due to the radiation pressure. If additionally $T_2=300 K$, the oscillating force $F_2^2$ is visible and it  dominates the total force for large $d$ if $T_1=0 K$; the net force now has many zero crossings, where every second one is a {\it stable equilibrium point}.  As discussed above, we expect the wavelength of the oscillations to be roughly $4.75$~$\mu$m due to the resonance of SiO$_2$ at wavelength $9.5$~$\mu$m. Additional modulations are due to interferences with a second resonance of SiO$_2$ at $22$~$\mu$m. The figure also provides complete information about the force on sphere~1: e.g., in case $T_1=0$ and $T_2=300K$, the red curve shows the force acting on sphere 2, while the green curve shows the force on sphere 1. At the crossing of the solid red and dashed green curves the two spheres feel equal forces in {\it the same direction}.  This corresponds to what we define as a self-propelled pair (SPP), where the spheres
experience equal acceleration in the same direction and hence remain at a fixed separation.
Note, however, that this is an unstable arrangement in which any small perturbation leads to the spheres moving apart.

Figure~\ref{fig:1a} shows the situation for a warm ($300 K$) environment. Here, the force has repulsive parts only if $T_1= 300 K$ and $T_2=0 K$ where it shows multiple stable equilibrium points. For all other cases, the force is purely attractive, decaying as $1/d^2$ if $T_1=0$. For $T_1=0$ and $T_2=300K$, one has stable and unstable SPP's, e.g., where the black dotted curve crosses the dashed green curve in Fig.~\ref{fig:1a}, i.e., for $R_2=R_1 2^{1/3}$ assuming solid spheres with mass $\sim R_j^3$.
\begin{figure}
\includegraphics[angle=270,width=0.9\linewidth]{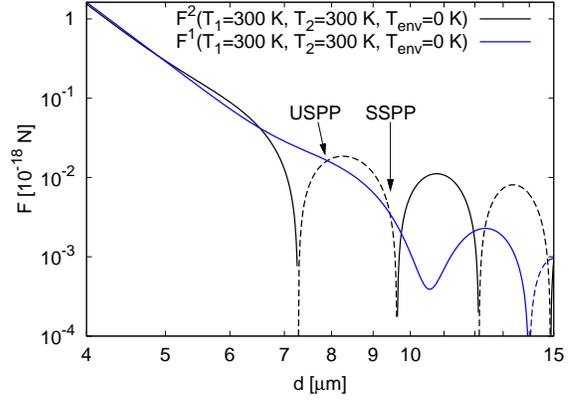}
\caption{\label{fig:2} Total forces on SiO$_2$ spheres with shifted optic resonances. Dashed lines indicate repulsion. Unstable (U) and stable (S) SPP and zero force points are visible.}
\end{figure}

The $1/d^2$ contribution to $F_1^2$ in Eq.~\eqref{eq:f1} (with Eq.\eqref{eq:expT}) is proportional to the product of the imaginary parts of the polarizabilities. These are peaked at the resonances of the material and this term can be suppressed by reducing the overlap of resonances. Figure~\ref{fig:2} shows the forces where the dielectric function of sphere 2 is replaced by $\tilde\varepsilon_2(\omega)=\varepsilon_{SiO_2}(1.17\omega)$, which, in principle,  can be achieved by using different isotopes.  Now, the forces are asymmetric even for $T_1=T_2$, and due to the suppression of $F_1^2$, we have stable  as well as unstable SPP's for e.g. $T_1=T_2=300K$, and $T_{env}=0K$, in contrast to Fig.~\ref{fig:1}.

For a sphere ($R$, $\varepsilon_s$, $\mu_s$, $T_s$) in front of a plate ($\varepsilon_p$, $\mu_p$, $T_p$) at center to surface separation $d$, Eq.~\eqref{eq:forcef} gives distinct non-equilibrium forces acting on the plate, or on the sphere. While both can be derived with equal effort, we restrict to the force acting on the sphere, ${F}^s=-{\bf F}^s\cdot {\bf n}_p$ (with outward normal ${\bf n}_p$ of the plate), separated into $F_p^s$ and $F_s^s$. 
Scattering from the plate is governed by the Fresnel reflection coefficients $r^P$ for  $P=M$, $N$, given by
\begin{align}
r^M\left(k_\perp,\omega\right)=\frac{\mu(\omega)\sqrt{\frac{\omega^2}{c^2}-k_\perp^2}-\sqrt{\varepsilon(\omega)\mu(\omega)\frac{\omega^2}{c^2}-k_\perp^2}}{\mu(\omega)\sqrt{\frac{\omega^2}{c^2}-k_\perp^2}+\sqrt{\varepsilon(\omega)\mu(\omega)\frac{\omega^2}{c^2}-k_\perp^2}}\notag,
\end{align}
with $r^N$ obtained from $r^M$ by interchanging $\mu$ and $\varepsilon$. 
In the one reflection approximation, the force $F_p^s$ is derived by a one-time-scattering of the radiation of the plate at the sphere. The subsequent integration in Eq.~\eqref{eq:fin} is done in plane waves basis, over two planes, enclosing the sphere and separating it from the plate. For the contribution $F_s^s$, the sphere radiation is scattered at the plate with identical surface of integration. 
As before, this procedure (valid for $d\gg R$)~\cite{Kruger11b} involves all multipoles of the sphere. 
Only for $R\ll\lambda_T$,  we can further restrict the $T$ operators to $l=1$.
The interaction term $F_p^s(T)$ has two distinct contributions,  $F_p^s=F_{p,pr}^s+F_{p,ev}^s$, from propagating and evanescent waves emitted by the plate, 
\begin{align}
&\lim_{\{d,\lambda_T\}\gg R}F_p^s=\frac{3\hbar}{2c\pi}\int_0^\infty d\omega\frac{\omega}{e^{\frac{\hbar\omega}{k_BT}}-1}\left(f_{pr}+f_{ev}\right)\label{eq:spf},
\end{align}
where the functions 
\begin{align}
f_{pr}&=\left(\frac{c}{\omega}\right)^2\int_{0}^{\omega/c}k_\perp dk_\perp\sum_{P,P'}(1- |r^P|^2)\Re[{\cal T}^{P'}],\\
\notag f_{ev}&=2\left(\frac{c}{\omega}\right)^2\int_{\omega/c}^{\infty}k_\perp dk_\perp e^{-2d\sqrt{k_\perp^2-\omega^2/c^2}}\\&\sum_{P}\Im\left[r^P\left(2 \frac{k^2_\perp c^2}{\omega^2}-1\right)+r^{\bar{P}}\right]\Im[{\cal T}^P],
\end{align}
explicitly contain the radiation of the plate \cite{Rytov3}. The force $F_{p,pr}^s$ is $d$ independent as it arises from absorption or scattering of far field photons by the sphere, while the near field contribution 
$F_{p,ev}^s$ depends on $d$.
The self term $F_s^s(T)$
\begin{align}
&\lim_{\{d,\lambda_T\}\gg R}F_{s}^s\notag=\frac{-3\hbar c}{\pi}\sum_{P}\int_0^\infty d\omega\frac{\Re[{\cal T}^P]}{\omega(e^{\frac{\hbar\omega}{k_BT}}-1)}\int_{0}^{\infty}k_\perp dk_\perp \\&\Re\left\{e^{2id\sqrt{\omega^2/c^2 -k_\perp^2}}\left[r^P\left(2 \frac{k^2_\perp c^2}{\omega^2}-1\right)+r^{\bar{P}}\right]\right\},\label{eq:Fss}
\end{align}
contains both evanescent and propagating contributions but no separation independent term. Instead, $F_{s}^s$ behaves similarly as $F_{2}^2$ in Eq.~\eqref{eq:selft}, oscillating as a function of $d$, falling off at large separations as $1/d$.

For a dielectric sphere and plate, we next employ Eqs.~\eqref{eq:insu} and \eqref{eq:al},
to obtain the leading behavior at low temperatures ($\lambda_T\gg\{\lambda_0,R\}$, but not necessarily $d$). 
The $d$ independent part now becomes,
\begin{equation} 
\lim_{d\gg R}F_{p,pr}^s=-\frac{8\pi^5}{63} \frac{\hbar c}{\lambda_T^6}f_{pr}(\omega=0) \lambda_{in,s}\alpha_{i0}.\label{eq:propf}
\end{equation} 
$F_{p,ev}^s$ can be analyzed in the following two limits, corresponding to expansions of the function $f_{ev}(\omega,d)$,   
\begin{align}
\lim_{d\gg\lambda_T\gg \{R,\lambda_0\}} F_{p,ev}^s=\frac{\pi}{6}\frac{\hbar c }{\lambda_T^2d^3} \Re\left[\frac{1+\varepsilon_{0,p}}{\sqrt{\varepsilon_{0,p}-1}}\right] \alpha_{0}.\label{eq:evf}
\end{align}
In the opposite limit, with $\lambda_T\gg\{d,R_j,\lambda_0\}$, we have
\begin{align}
\lim_{d\gg R}F_{p,ev}^s=\frac{\pi}{2}\frac{\hbar c \lambda_{in,p}}{\lambda_T^2d^4} \frac{1}{(1+\varepsilon_{0,p})^2} \alpha_{0}.\label{eq:evc}
\end{align}
Equation~\eqref{eq:evf} is similar to Eq.~(12) in Ref.~\cite{Antezza05}.
As was the case for $F_2^2$, in leading order in temperature the self part $F_s^s$ does not oscillate. For $d\gg\lambda_T\gg \{R,\lambda_0\}$, we have $F_s^s\propto 1/d^6$, the counterpart of Eq.~\eqref{eq:larged}, with a lengthy prefactor. For $\lambda_T\gg\{d,R_j,\lambda_0\}$ we have
\begin{equation}
\lim_{d\gg R}F_{s}^s=\frac{\pi}{4}\frac{\hbar c}{\lambda_T^2 d^4}\frac{\varepsilon_{0,p}-1}{\varepsilon_{0,p}+1} \lambda_{in,s}\alpha_{i0},
\end{equation}
which is identical to Eq.~\eqref{eq:evc} when interchanging real and imaginary parts for $r^P$ and $\alpha$. 
\begin{figure}
\includegraphics[angle=270,width=0.9\linewidth]{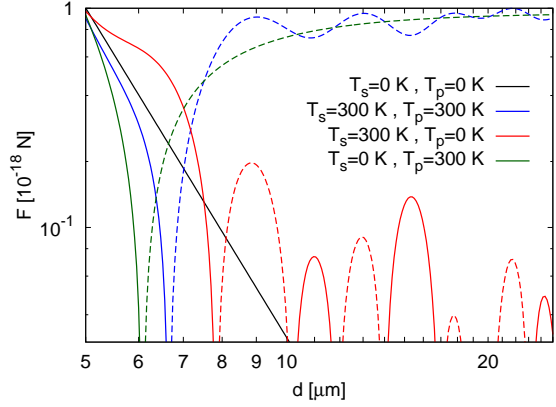}
\caption{\label{fig:3} Total force on a SiO$_2$ sphere of $R=1\mu$m in front of a SiO$_2$ plate in a cold environment. Dashed lines indicate repulsion. Every second zero of the red curve is a stable equilibrium point.}
\end{figure}
\begin{figure}
\includegraphics[angle=270,width=0.9\linewidth]{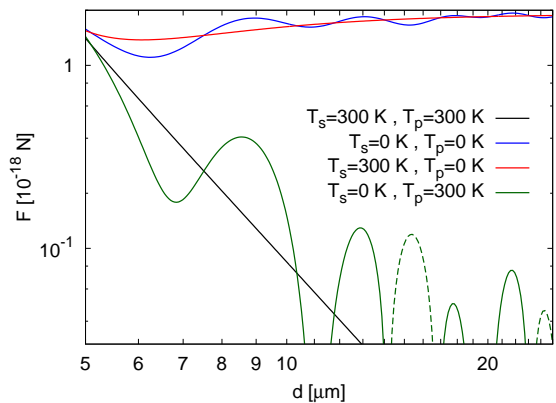}
\caption{\label{fig:3a} Total force on a SiO$_2$-sphere of $R=1\mu$m in front of a SiO$_2$ plate in a warm ($300 K$) environment. Dashed lines represent repulsion.}
\end{figure}
The equilibrium force can be found in Ref.~\cite{Antezza04}. For $d\gg R$ one has
\begin{subequations}
\begin{align}
\lim\limits_{\lambda_T\gg d\gg R}F^{s,eq}&=\frac{3}{2\pi}\frac{\hbar c }{d^5}\frac{\varepsilon_{0,p}-1}{\varepsilon_{0,p}+1}\alpha_{0}\Phi(\varepsilon_{0,p}), \\
\lim\limits_{d\gg \{R,\lambda_T\}}F^{s,eq}&=\frac{3\hbar c }{4d^4\lambda_T}\frac{\varepsilon_{0,p}-1}{\varepsilon_{0,p}+1}\alpha_{0},
\end{align}
\end{subequations}
where $\Phi(\varepsilon_{0,p})$ is e.g. given in Ref.~\cite{Antezza04}.  Figure~\ref{fig:3} shows numerical results for the force on a sphere in front a plate (both made of SiO$_2$) for $R=1\mu$m in a cold (0 $K$) environment. Again, we use the simplification of Eq.~\eqref{eq:expT} and the resulting force is proportional to $R^3$ (also here, $R=1\mu$m is roughly the upper bound of validity of this simplification). If the plate is warm, the distance independent repulsion is visible. If only $T_s$ is different from $T_{env}$, the force $F_s^s$ dominates at large $d$, leading to multiple stable points. 

Figure~\ref{fig:3a} shows the curves for a warm ($300 K$) environment. Here, the $d$ independent force (for $T_p=0$) is attractive. Again, if  only $T_s$ is different from $T_{env}$, we observe many changes of the sign of the force. 
\begin{figure}
\includegraphics[angle=270,width=0.9\linewidth]{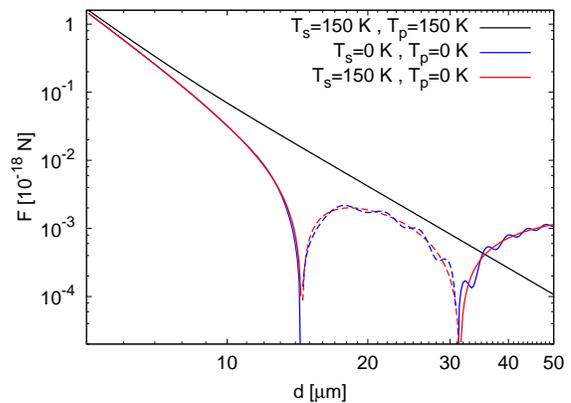}
\caption{\label{fig:4} Total force on a SiC sphere in front a SiO$_2$ plate in an environment at 150 K. Dashed lines indicate repulsion. The second zero is a stable equilibrium point. Eqs.~\eqref{eq:spf} and \eqref{eq:Fss}, together with Eq.~\eqref{eq:expT}, are strictly valid for SiC spheres with $R\lesssim0.3\mu$m, but for comparison to the previous graphs we show the force  computed for $R=1\mu$m.}
\end{figure}
Exploring the effects of shifting resonances, we found that in contrast to the case of two spheres, here shifting suppresses the self term more strongly than the interaction term. For the special case of a (resonance-shifted) SiC sphere\footnote{We found that the described effect appears most pronounced by adjusting the resonance of SiC (see Ref.~\cite{Spitzer59} for $\varepsilon(\omega)$) by insertion of a factor of 0.75 in frequency.} in front of a SiO$_2$ plate, see Fig.~\ref{fig:4}, the temperature of the sphere is almost irrelevant for the force. 
This is beneficial to experimental setups, as it is presumably harder to maintain $T_s$ at a constant value, compared to keeping $T_p$ and $T_{env}$ constant. Additionally, in Fig.~\ref{fig:4}, the special choice of parameters leads to a stable equilibrium point, which, again, is almost independent of $T_s$.  

The presented formulae for the forces, i.e., Eqs.~\eqref{eq:f1}, \eqref{eq:selft}, \eqref{eq:spf} and \eqref{eq:Fss} after substitution of $T_{j}^{N}=i\frac{2\omega^3}{3c^3} \alpha_{j}(\omega)$ and $T_{j}^{M}=i\frac{2\omega^3}{3c^3} \beta_j(\omega)$ with $\beta_j(\omega)=((\mu_j(\omega)-1)/(\mu_j(\omega)+2))R_j^3$ the magnetic dipole polarizability, can be derived independently from diluting two plates of vanishing thickness, confirming the correctness of our formalism for compact objects. This calculation will be presented elsewhere. 

Our results constitute a
macroscopic generalization of non-equilibrium interactions between thermal gases of atoms and interactions between atoms in excited states \footnote{There are intriguing similarities to previous work on atoms: Our interaction force in Eq.~\eqref{eq:spf} shares certain terms with the studies in Refs.~\cite{Henkel02,Antezza05}. The force in Ref.~\cite{Ellingsen10} shows similar behavior as our self force in Eq.~\eqref{eq:Fss}. Additionally, the last three terms of our Eq.~\eqref{eq:f1} have common structure as Eq.~(2.17) in Ref.~\cite{Power94}.}. However, we emphasize that the forces on two macroscopic objects are not equal and and opposite, an effect which cannot be found from the interaction potential as used in studies of two atoms \cite{Kweon93,Power94,Cohen03,Rodriguez10}. We hope that our results may eventually shed new light on the debated non-equilibrium interactions of atoms.

While, for simplicity, we discussed the forces for small radii and moderate temperatures, our formalism is more generally applicable for any values of $R$, $d$ and $T$. Future work will consider the cases of larger spheres where non-equilibrium effects may be stronger.

This research was supported by the DFG grant No. KR 3844/1-1, NSF Grant No.  DMR-08-03315 and DARPA contract No. S-000354.
We thank R.~L.~Jaffe, N.~Graham, M.~T.~H.~Reid, M.~F.~Maghrebi and V.~A.~Golyk for discussions.

\end{document}